\documentclass[11pt]{article}
\usepackage[letterpaper,margin=1in]{geometry}
\usepackage[T1]{fontenc}
\usepackage[utf8]{inputenc}
\usepackage{lmodern}
\usepackage{amsmath,amssymb,booktabs,array,graphicx}
\usepackage{xcolor,enumitem,microtype}
\usepackage{seqsplit}
\usepackage{float}
\newcommand{\hash}[1]{\texttt{\seqsplit{#1}}}
\usepackage[numbers,sort&compress]{natbib}
\usepackage{hyperref}
\usepackage{tikz}
\usetikzlibrary{arrows.meta,positioning}
\hypersetup{colorlinks=true,linkcolor=blue!45!black,citecolor=blue!45!black,urlcolor=blue!45!black,pdftitle={FIDES-Live: Evaluating Behavioral Concordance in Sequential Trading Agents},pdfauthor={Arther Tian, Alex Ding, Simon Wu, Aaron Chan}}
\newcommand{\CryptoN}{7,299}
\newcommand{\CryptoDecisionPass}{7,052}
\newcommand{\CryptoDecisionDen}{7,244}
\newcommand{\CryptoReportPass}{6,597}
\newcommand{\CryptoReportDen}{7,265}
\newcommand{\CryptoCommitDen}{3,065}
\newcommand{\CryptoCommitPass}{2,892}

\newcommand{\CryptoFills}{221}
\newcommand{\CryptoOrders}{338}
\newcommand{\CryptoDecisionRate}{97.35}
\newcommand{\CryptoReportRate}{90.81}
\newcommand{\CryptoCommitRate}{94.36}
\newcommand{\CryptoCoverageRate}{25.13}
\newcommand{\CryptoDecisionYield}{96.62}
\newcommand{\CryptoReportYield}{90.38}
\newcommand{\CryptoActiveRate}{3.11}
\newcommand{\EquityN}{310}
\newcommand{\EquityDecisionPass}{297}
\newcommand{\EquityDecisionDen}{309}
\newcommand{\EquityReportPass}{274}
\newcommand{\EquityReportDen}{309}

\newcommand{\EquityDecisionRate}{96.12}
\newcommand{\EquityReportRate}{88.67}
\newcommand{\EquityCommitRate}{97.03}

\newcommand{\AllRounds}{7,609}
\newcommand{\CryptoCalls}{14,833}
\newcommand{\TokenTotal}{211,324,566}
\newcommand{\CryptoMacroDecisionRate}{97.34}
\newcommand{\CryptoMacroReportRate}{89.66}
\newcommand{\AuditControlCount}{36}

\title{\textbf{FIDES-Live: Evaluating Behavioral Concordance\\in Sequential Trading Agents}}
\author{Arther Tian$^{a}$, Alex Ding$^{a,*}$, Simon Wu$^{a}$, Aaron Chan$^{a}$\\[4pt]
\small $^{a}$DGrid AI\\
\small $^{*}$Corresponding author: \texttt{alex.ding@dgrid.ai}}
\date{}
\begin{document}
\maketitle
\begin{abstract}
Sequential trading agents produce decisions, orders, execution receipts, account states, and reports that need not agree, even when the final portfolio value is correctly computed. We present FIDES-Live, an operational extension of FIDES from generated strategy concordance to the audit of sequential agent traces. The protocol separates structured decision validity, optional numerical commitment adherence, execution and ledger reconciliation, and receipt-grounded reporting, with explicit coverage and unscorable outcomes. We retrospectively audit ten requested inference routes over \CryptoN{} cryptocurrency account-rounds and a \EquityN{}-round equity pilot, using real market observations and local paper execution. Cryptocurrency decision validity is \CryptoDecisionRate\% and structured reporting accuracy is \CryptoReportRate\% among scorable responses, but route-level decision and reporting performance can diverge sharply. Commitment adherence is \CryptoCommitRate\% overall; it falls to 62.63\% on constraints emitted with nonempty order proposals, compared with 97.56\% for valid empty proposals. All retained cash and position transitions reconcile with recorded fills, while execution controls and unavailable quotes explain separate rejection populations. Short-window marked returns provide descriptive context rather than a new test of whether concordance predicts alpha. These observations motivate reporting concordance, evidence coverage, activity, and financial outcomes separately. They do not establish long-run trading skill, semantic faithfulness of explanations, or causal benefits from audit feedback.
\end{abstract}

\section{Introduction}
An investment agent can submit a valid order, fail to execute it, and subsequently describe it as completed. It can maintain an internally consistent account while violating its own allocation commitment. It can also report every transaction correctly and still lose money. These cases concern different properties, yet a portfolio return or a single aggregate score can make them indistinguishable. As language models move from proposing strategies to repeatedly acting on market observations, evaluating the artifacts surrounding each action becomes a separate measurement problem.

Our earlier FIDES protocol examined concordance among a natural-language strategy, its generated implementation, and its backtested outcomes~\cite{tian2026fides}. The present study follows that work but changes the measurement object. Instead of a strategy generated once and subsequently executed, we consider an agent that observes a changing market and its own holdings, proposes orders, encounters execution controls, and reports the resulting state. A discrepancy can originate in the proposal, the environment, execution, or reporting. Assigning all unsuccessful interactions to the model obscures this distinction; assigning every correct final balance to model competence obscures it in the other direction.

Live and sequential financial evaluation is already an established research setting~\cite{chen2025stockbench,qian2025whenagents,yu2025livetradebench}. Our contribution is a narrower operational audit. FIDES-Live reconstructs a trace linking observations, decisions, explicit commitments, proposed orders, receipts, account states, and reports. Deterministic checks test the properties supported by those artifacts. Decision validity tests the action contract. Commitment checks test numerical constraints the model actually states. Ledger reconciliation checks arithmetic against recorded fills. Reporting checks compare structured factual claims with the receipt provided to the model. Missing evidence and explicitly unavailable calls remain visible rather than receiving a convenient score.

We apply the protocol retrospectively to two forward-running paper-trading environments: a cryptocurrency spot environment with ten requested model routes and an equity pilot using the same roster. The retained corpus comprises \AllRounds{} account-rounds: \CryptoN{} cryptocurrency observations across eight UTC dates and \EquityN{} equity observations across two trading days. All market interactions use real public data, but fills are produced by local paper executors. The environments differ in their risk controls, advisory signals, costs, and scheduling; we analyze them separately. Chronological account-rounds are dependent observations, not thousands of independent market experiments.

Three results motivate the audit. First, valid decisions and accurate reports separate sharply for some inference routes: in cryptocurrency traces, the DeepSeek route has 98.73\% decision validity but 14.33\% structured reporting accuracy among scorable responses. This is a finding about the recorded route and requested format, not a claim about immutable model weights or financial reasoning. Second, optional commitment adherence is difficult to interpret without coverage and activity. Cryptocurrency adherence is \CryptoCommitRate\% overall, but 62.63\% on constraints emitted with nonempty proposals and 97.56\% with valid empty proposals. Third, execution failures and accounting errors do not coincide. All retained cash and position transitions reconcile with recorded fills, although the equity pilot contains 100 orders rejected for market unavailability and the cryptocurrency executor rejects 117 orders through its proposal gate.

\paragraph{Contributions.}
\begin{itemize}[leftmargin=*,itemsep=2pt]
\item An operational concordance protocol for sequential trading traces, with explicit eligibility, commitment coverage, and separate decision, execution, and reporting measurements.
\item A provenance-tracked retrospective audit of two forward-running environments, with reproducible aggregation from normalized outcomes, file-integrity checks, timestamp-based sequence reconstruction, and conservative treatment of unavailable valuation evidence.
\item An empirical decomposition showing how report format failures, optional commitments, activity, and execution controls change the interpretation of aggregate reliability measures.
\end{itemize}
The study establishes measurements and observed discrepancies. It does not establish profitable trading, internal reasoning faithfulness, deliberate deception, or a causal benefit from providing audit feedback.

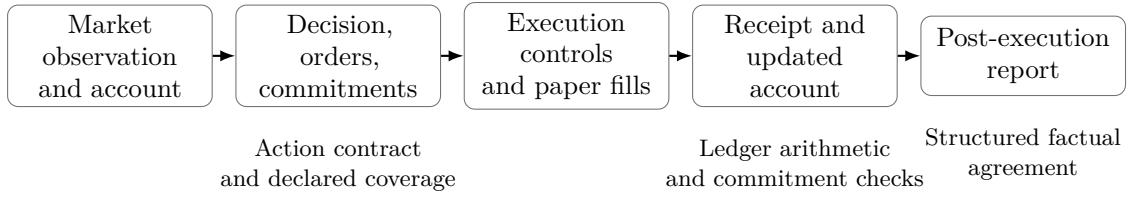
\begin{figure}[t]
\centering
\begin{tikzpicture}[node distance=3mm,box/.style={draw=black!55,rounded corners,align=center,font=\small,text width=2.45cm,minimum height=1.05cm},arr/.style={-{Latex[length=2mm]},thick}]
\node[box] (obs) {Market observation\\and account};
\node[box,right=of obs] (dec) {Decision, orders,\\commitments};
\node[box,right=of dec] (exe) {Execution controls\\and paper fills};
\node[box,right=of exe] (rec) {Receipt and\\updated account};
\node[box,right=of rec] (rep) {Post-execution\\report};
\draw[arr] (obs)--(dec);\draw[arr] (dec)--(exe);\draw[arr] (exe)--(rec);\draw[arr] (rec)--(rep);
\node[font=\footnotesize,align=center,below=3mm of dec] {Action contract\\and declared coverage};
\node[font=\footnotesize,align=center,below=3mm of rec] {Ledger arithmetic\\and commitment checks};
\node[font=\footnotesize,align=center,below=3mm of rep] {Structured factual\\agreement};
\end{tikzpicture}
\caption{The FIDES-Live measurement object. Intended orders, execution outcomes, and reported facts are distinct artifacts. The deterministic audit checks supported properties and preserves unavailable evidence; it does not infer hidden reasoning from a correct account balance.}
\label{fig:protocol}
\end{figure}

\section{Related Work}
\label{sec:related}

\paragraph{Trading agents and live evaluation.}
Financial agent research increasingly evaluates actions and portfolios rather than isolated financial answers. TradingAgents organizes analysis, deliberation, trading, and risk management into specialized roles~\cite{xiao2024tradingagents}. StockBench studies sequential stock decisions with market, fundamental, and news inputs~\cite{chen2025stockbench}. Agent Market Arena standardizes live inputs and execution feedback across agent architectures in equity and cryptocurrency environments~\cite{qian2025whenagents}. LiveTradeBench extends live evaluation to portfolio allocation across stock and prediction markets~\cite{yu2025livetradebench}. Thus, live data, sequential decisions, and multiple markets are established evaluation settings. AlphaForgeBench further diagnoses unstable action sequences and instead evaluates generated factors and strategy implementations through deterministic execution~\cite{zhang2026alphaforgebench}. FIDES-Live retains sequential agent actions and examines agreement among the artifacts surrounding them. Its purpose is to complement financial outcome evaluation with an inspectable operational audit, rather than establish a new claim to live-market evaluation.

\paragraph{State-grounded reliability and explanation faithfulness.}
AgentBench evaluates reasoning and action across interactive environments, including the failure modes arising during interaction~\cite{liu2023agentbench}. The state-based evaluation in $\tau$-bench compares final database states with annotated goals and measures reliability across repeated trials~\cite{yao2024taubench}. Our audit similarly uses observable state as evidence, but consecutive trading rounds are not independent repetitions of one task; they cannot support a direct interpretation as $\mathrm{pass}^{k}$. A separate literature shows that plausible explanations can omit factors influencing model outputs~\cite{turpin2023unfaithful}. Agreement with a ledger is a narrower property: it establishes whether checkable claims match observable records, not whether a narrative reveals the model's internal causal reasoning. We therefore use \emph{concordance} for artifact agreement and avoid interpreting detected discrepancies as evidence of deliberate deception.

\paragraph{From FIDES to sequential operational concordance.}
FIDES reconciles natural-language strategy descriptions, generated code, and backtested outcomes through three concordance dimensions~\cite{tian2026fides}. FIDES-Live extends the measurement object from one generated strategy to a sequence of decisions, submitted orders, execution receipts, account states, and reports. This changes the attribution problem: a valid decision may encounter unavailable quotes or rejected execution, while an inaccurate report may incorrectly describe what subsequently occurred. We keep these events distinguishable. Moreover, factual accuracy after observing a receipt does not validate an earlier investment forecast. The numerical checks reported here assess the structured commitments and factual fields supported by the retained traces; they do not constitute a complete semantic evaluation of every rationale.

\paragraph{Measurement before aggregation.}
Our earlier PoQ work examines evaluator robustness, dimension reliability, and evaluation cost~\cite{tian2026robustpoq,tian2026multidimensional,tian2026poqjudge}. SFGA applies a related principle: inexpensive measurements should be used within their evidential limits, with uncertainty made explicit~\cite{tian2026sfga}. FIDES-Live follows this measurement philosophy by prioritizing deterministic, inspectable checks and reporting eligible denominators and evidence coverage. These connections motivate design choices; they do not imply that the present corpus was produced with PoQ consensus, a trained PoQ-Judge, or SFGA adjudication. We report operational dimensions separately rather than treating an unvalidated weighted composite as an established measure of overall agent quality.

\paragraph{Relationship to the preceding FIDES study.}
Table~\ref{tab:prior-comparison} makes the change in measurement explicit.
The preceding study evaluates 40 generated strategy outputs using historical
backtests~\cite{tian2026fides}; the present corpus contains newly recorded
forward-running account traces and does not reuse those strategy outputs as
observations. This is a continuation of the measurement perspective, rather
than a replication of its three scores. Action-contract validity does not
replace the earlier language--code comparison, and accurate reporting after
seeing a receipt does not validate an earlier forecast. Consequently, scores
from the two protocols are not directly comparable. The separation between
concordance and financial performance is inherited motivation, not a new
claim of statistical independence or profitable trading. The new empirical
focus is the attribution of discrepancies across sequential proposals,
execution controls, marked states, and structured reports, with explicit
eligibility and coverage. Neither a semantic explanation benchmark nor an
estimate of financial alpha is supplied by these measurements.

\begin{table}[t]
\centering\small
\begin{tabular}{>{\raggedright\arraybackslash}p{0.21\textwidth}>{\raggedright\arraybackslash}p{0.35\textwidth}>{\raggedright\arraybackslash}p{0.35\textwidth}}
\toprule
Property & Earlier FIDES & FIDES-Live \\
\midrule
Observation unit & One generated strategy; 40 outputs & One retained account-round; \AllRounds{} rounds \\
Evidence & Strategy description, generated code, historical backtest & Decision, commitments, order receipts, account state, report \\
Measurements & Language--code, implementation, and claim--outcome concordance & Action-contract validity, optional commitments, ledger reconciliation, structured report agreement \\
Financial setting & Historical out-of-sample strategy evaluation & Forward-running local paper execution on real market observations \\
\bottomrule
\end{tabular}
\caption{The studies share an audit perspective but use different observations and measurement definitions. Their scores are not directly comparable.}
\label{tab:prior-comparison}
\end{table}

\section{An Operational Concordance Protocol}
\label{sec:methods}

\subsection{The trace as the measurement object}
An agent round produces several artifacts that should describe one interaction but need not agree. For model $m$ and round $t$, write
\begin{equation}
\tau_{m,t}=(o_{m,t},s^-_{m,t},d_{m,t},u_{m,t},e_{m,t},s^+_{m,t},r_{m,t}),
\end{equation}
where $o_{m,t}$ is the supplied market observation, $s^-$ the pre-execution account, $d$ the decision response, $u$ its proposed orders, $e$ the execution receipts, $s^+$ the post-execution account, and $r$ the subsequent report. The observation can differ across accounts, even within a scheduled interval. The decision can contain optional numerical commitments. Orders and receipts are distinct: an intended buy can be rejected and leave the account unchanged. The report observes the receipt and cannot place further orders. Its accuracy is therefore a post-execution factual property, not a test of an earlier investment prediction.

The unit of analysis is an \emph{account-round}, not an API call or an independent market trial. Accounts maintain separate cash and holdings, while observations and market shocks can be shared. We evaluate the equity and cryptocurrency environments separately. Differences in their inputs and execution controls prevent interpreting pooled counts as one homogeneous treatment.

\subsection{Evidence reconstruction and eligibility}
We recompute measurements offline from retained round artifacts rather than reuse dashboard scores. The reconstruction records each source's SHA-256 digest and preserves its decision response, receipts, account states, report, timestamps, and available call metadata. Indexed cryptocurrency round artifacts are checked against their recorded hashes and sizes. Summaries are not substitutes for full artifacts. Synthetic demonstrations and historical replay runs are excluded. Per-account sequences are ordered by recorded event timestamps; an ordinal that reverses the temporal sequence is a provenance anomaly to disclose, not a reason to silently rewrite its contents.

Each applicable check has three outcomes: pass, fail, or unscorable. An available response that violates the requested schema is a failure. Explicit \texttt{api\_error}, \texttt{agent\_error}, and \texttt{interrupted} statuses are treated as call unavailability and excluded from conditional content denominators; these labels do not identify which upstream component caused the failure. An unattempted report has no report-content outcome. We retain unavailable and unattempted cases in accounting totals. Execution rejection is separately classified from decision validity and report accuracy.

Let $N$ be retained account-rounds, $\mathcal{D}$ the rounds with scorable decision responses, and $\mathcal{R}$ those with scorable attempted reports. With pass indicators $v_i$ and $a_i$, respectively, we report
\begin{equation}
V=\frac{\sum_{i\in\mathcal{D}}v_i}{|\mathcal{D}|},\qquad
A=\frac{\sum_{i\in\mathcal{R}}a_i}{|\mathcal{R}|}.
\end{equation}
We also give unavailable counts and successful outputs divided by $N$, termed \emph{retained-round yield}. These quantities describe availability-adjusted output success within committed retained rounds, not completion probabilities across every started execution attempt. Attempts that never produce a retained round are outside this denominator. Primary rates pool eligible observations (micro-averaging); equal-route macro-averages give each of the ten requested routes one weight. Empty denominators remain undefined. Neither unavailable evidence nor an absent optional commitment receives automatic credit.

\subsection{Decision validity and explicit commitments}
A valid decision is a JSON object containing a nonempty \texttt{rationale}, an \texttt{orders} array, and at most an optional \texttt{commitments} object. Each order has exactly an allowed symbol, a buy/sell side, and a finite positive quantity; its count must respect the recorded configuration. An empty array is valid. Commitments admit only finite fractions in $[0,1]$ for \texttt{min\_cash\_ratio} and \texttt{max\_position\_ratio}. Additional keys, malformed types, and nonfinite values fail validation. A syntactically valid order can still be unaffordable or rejected. Consequently, $V$ measures structured decision validity, not investment quality or semantic agreement between rationale and action. Order totals count orders in contract-valid decisions only; malformed or unavailable decisions are neither valid empty proposals nor admitted nonempty proposals. Recognizable order objects inside an invalid response do not enter the admitted-order population.

For a valid decision, let $\mathcal{C}^{\mathrm{decl}}_i$ contain its declared nontrivial commitments and $\mathcal{C}_i\subseteq\mathcal{C}^{\mathrm{decl}}_i$ the evaluable subset. Given post-execution cash $c_i$, equity $E_i>0$, quantities $q_{i,s}$, and execution-midpoint prices $p_{i,s}$, define $w_{i,s}=q_{i,s}p_{i,s}/E_i$. Stored weights are cross-checked against this construction wherever complete price evidence is available. The two supported checks are
\begin{align}
c_i/E_i+\epsilon &\geq \ell_i &&\text{(minimum cash)},\\
\max\bigl(\{w_{i,s}:s\in\mathrm{holdings}_i\}\cup\{0\}\bigr) &\leq h_i+\epsilon &&\text{(maximum position)},
\end{align}
with $\epsilon=10^{-9}$. They refer to the recorded post-execution marked state, including actual rejections, rather than a hypothetical full fill or future return. A zero cash floor and a unit position ceiling follow from the long-only, unlevered rules; we count them separately as trivial and exclude them from substantive compliance.

Commitments are evaluable only when the state has finite cash, positive finite equity, and finite position weights, with no recorded market/execution error or stale-state flag. Each held asset must have a finite positive uncrossed bid--ask quote, spread no greater than 100 basis points, and a timestamp between zero and 90 seconds before the execution snapshot in cryptocurrency, or 120 seconds in equities. These are project-defined eligibility thresholds, not universal financial benchmark standards. Otherwise the declared constraint is unscorable; a stale mark does not establish either compliance or violation.

Let $\mathcal{V}$ be valid retained decisions. With $z_{i,c}$ indicating that constraint $c$ passes, commitment coverage and compliance are
\begin{equation}
\mathrm{Coverage}=\frac{\sum_{i\in\mathcal{V}}\mathbf{1}[|\mathcal{C}^{\mathrm{decl}}_i|>0]}{|\mathcal{V}|},\qquad
\mathrm{Compliance}=\frac{\sum_{i\in\mathcal{V}}\sum_{c\in\mathcal{C}_i}z_{i,c}}{\sum_{i\in\mathcal{V}}|\mathcal{C}_i|}.
\end{equation}
Coverage includes declarations whose marked state is unscorable; the difference $\sum_i(|\mathcal{C}^{\mathrm{decl}}_i|-|\mathcal{C}_i|)$ is reported as unknown. The two measures answer different questions. High compliance at low coverage is weak evidence about general adherence; emitting no commitment is not a successful adherence test. Platform-imposed risk limits are not voluntary model commitments.

\subsection{Receipt-grounded reporting}
A report must contain exactly \texttt{summary}, \texttt{cash}, \texttt{equity}, \texttt{positions}, and \texttt{filled\_orders}. The summary must be a nonempty string. Cash and equity must agree with the supplied post-execution account within an absolute tolerance of $0.01$ monetary units. The positions map must have exactly the recorded symbol set, with every quantity within $10^{-6}$ asset units. The filled-order count must be an integer, excluding Boolean values, equal to the number of receipts marked filled \emph{in that round}. A report passes only when its structure and all four factual fields pass. Component failures can overlap and are reported separately.

These checks evaluate structured claims against the receipt even when its market valuation is stale. Receipt agreement and current-market valuation quality are distinct properties. The checks do not semantically score the summary, compare the rationale's prose with orders, infer hidden reasoning, or determine whether an error was intentional. No language-model judge or human semantic annotation is used for the reported measurements. A correct structured report can therefore coexist with an unchecked narrative inconsistency.

\subsection{Execution and ledger reconciliation}
We distinguish proposed, filled, and rejected orders and retain recorded rejection reasons and available pre-execution gate checks. A gate can block the whole proposal; an absence of fills is consequently not, by itself, evidence of inactivity or a model error. Gate-pass rates and filled/proposed counts are execution descriptors, not additional components of $V$ or $A$.

To check the state anchor independently of legacy evaluation flags, we reconstruct cash and positions from the pre-execution account and recorded filled orders. For filled order $j$, let $\sigma_j=1$ for a buy and $-1$ for a sell, $n_j$ its recorded notional, $q_j$ its quantity, and $f_j$ its fee. The reconstructed state satisfies
\begin{align}
\widehat c_i^+&=c_i^- -\sum_j(\sigma_j n_j+f_j),\\
\widehat q_{i,s}^+&=q_{i,s}^-+\sum_{j:\,s_j=s}\sigma_j q_j.
\end{align}
We compare reconstructed cash with the recorded state within $10^{-5}$ monetary units and quantities within $10^{-6}$ asset units, discarding position dust of magnitude at most $10^{-9}$. Where prices are available, we also check $E_i=c_i+\sum_s q_{i,s}p_{i,s}$ within $10^{-5}$ and reconstructed position weights within $10^{-9}$. Recorded account prices take precedence; execution bid--ask midpoints provide a fallback. Reconciliation establishes internal accounting agreement with recorded fills and marks. It does not attest exchange execution, quote freshness, or realistic liquidity. Returns and token usage remain separate descriptive outcomes; we introduce no weighted overall quality score.

For the settled session, marked return is $R_m=E^{\mathrm{end}}_m/E^{\mathrm{initial}}_m-1$ and final cash fraction is $c^{\mathrm{end}}_m/E^{\mathrm{end}}_m$. Initial allocations and the common eligible saved endpoint determine the return; no completeness of the intervening valuation path is assumed.

\section{Corpus and Experimental Setting}
\label{sec:setting}
\subsection{Two forward-running environments}
\paragraph{Cryptocurrency spot.}
The main corpus records decisions from 2026-09-30 03:44:09 to 2026-10-07 22:14:24 UTC. The universe is BTC, ETH, SOL, BNB, DOGE, and XRP. Public Binance spot bid--ask snapshots supply observations and execution references. Quotes are denominated in USDT; account displays assume one USDT equals one USD. This assumption ignores stablecoin basis risk. Snapshot timestamps identify HTTP snapshots, not independently verified exchange event times. Each model begins each seven-day session with a separate \$10,000 paper account. One session is settled; the following session is only partially observed. We retain 6,608 account-rounds from the first and 691 from the second, with no attempt to impute the unfinished period.

The intended decision interval is 900 seconds, with account marks approximately every 30 seconds. Agents receive up to 90 daily historical bars per asset, recent account history, and a deterministic momentum/volatility analysis. The analysis is advisory rather than an LSTM or other trained forecasting model. The recorded method combines 20-period momentum and the observed spot move, scaled by 20-period return volatility. A fixed balanced policy requires at least 20\% cash, at most 30\% in one asset, and at most 80\% total asset exposure. These limits belong to the platform policy, not to voluntary model commitments.

A pre-execution gate previews the entire proposal and checks schema, quotes, affordability, holdings, and risk limits. A failed check rejects the whole proposal without resizing individual orders. Even an empty proposal can fail when existing holdings drift beyond a risk limit. Long-only fractional quantities are allowed; shorting, borrowing, leverage, and external tools are unavailable. Filled buys use the ask and filled sells the bid, with 2 basis points of adverse slippage and a 10-basis-point notional fee. The maximum proposal contains 12 orders. Quote age is limited to 90 seconds and spread to 100 basis points. These are configured experimental assumptions, not universal benchmark standards or exchange fee tiers.

\paragraph{Equity pilot.}
The equity traces cover September 28--29, 2026. One empty initialization contains no account-rounds and is excluded from outcome denominators; two populated runs contain 5 and 26 joint decision rounds, respectively, or \EquityN{} account-rounds. The first uses SPY, QQQ, IWM, EFA, TLT, GLD, XLF, and XLE at a five-minute decision interval. The second adds AAPL, MSFT, NVDA, AMZN, GOOGL, META, and TSLA and uses 15-minute decisions during the US regular session, with 30-second marks. Alpaca IEX data provide bid--ask references. Each populated run starts independent \$10,000 model accounts with up to 30 historical daily bars. This operational evolution is retained rather than described as one frozen equity treatment.

The equity system supplies no mandatory balanced allocation gate or advisory momentum analysis. Its executor enforces cash and long-only holdings, charges 1 basis point in fees and 2 basis points of adverse slippage, admits at most 20 orders, and uses a 120-second quote-age limit; the main run records a 100-basis-point spread limit. In the observed unavailable-market rounds, the quote bundle is unavailable and the proposed orders receive explicit rejection receipts. A joint round shares an observation across models and waits for the decision batch before retrieving execution quotes; the cryptocurrency system instead records account-specific observation and execution snapshots. These differences rule out interpreting the two environments as a controlled test of market transfer.

\subsection{Inference routes and requested output}
The ten requested routes are GPT-4.1 mini, GPT-5.4, DeepSeek V3.2, Gemini 2.5 Flash, Claude Haiku 4.5, Qwen3 235B A22B Instruct 2507, Kimi K2.5, GLM-5.3, MiniMax M2.5, and Grok 4.20 non-reasoning. Table~\ref{tab:routes} supplies exact request identifiers. They are accessed through a compatible chat-completion gateway. Returned identifiers are retained, but do not attest fixed model weights; observed aliases include multiple DeepSeek and Claude identifiers. Consequently, model names in tables abbreviate requested routes.

Each account-round normally involves one decision request followed by one report request. Prompts permit empty orders and require a nonempty rationale; optional numerical commitments are explicitly distinguished from future-return predictions. The report receives the execution receipt and account state. No hidden repair call is added during this audit. The recorded maximum output allowance is 8,192 tokens with a 120-second timeout for cryptocurrency and 4,096 tokens with a 90-second timeout for the main equity run. No fixed sampling seed or weight-level version attestation is established by the retained evidence.

\subsection{Data reconstruction and validation}
The cryptocurrency database indexes \CryptoN{} round artifacts and 14,834 call artifacts. Every indexed round and call file is present and matches its recorded SHA-256 digest and byte length. The call table contains \CryptoCalls{} entries: 14,831 completed lifecycle records and two interrupted records. One additional artifact is an orphan started report, retained as provenance evidence but excluded from completed-call totals. A completed lifecycle is not necessarily a successful model response. The call population also includes attempts not represented by a committed round; it is not forced into twice the round count.

The database lists 840 unavailable ancillary attempt journals. Their associated committed round evidence is present, so they do not represent 840 missing executions. They do limit complete reconstruction outside committed traces. Historical compact database rows are expanded from full artifacts, which contain the original observations, decisions, receipts, and reports. The analysis excludes demonstration and historical replay runs, copied equity archives, and personal-use fixtures.

One recovered cryptocurrency record has a round ordinal inconsistent with its timestamp and original call payload. Sorting by recorded time restores cash and position continuity, so we retain the observation and disclose the anomaly. There are no cash or position discontinuities across 7,279 consecutive cryptocurrency account transitions or 290 equity transitions after this normalization. Hashes establish consistency with the stored index, not external notarization. All reported checks are recomputed offline; legacy decision-validity and report-accuracy flags agree with the recomputation, whereas conservative valuation eligibility removes 55 commitment checks from the compliance denominator.

We additionally reparse all retained no-error completed cryptocurrency call responses (14,034) and no-error equity participant responses (584) using the runtime's JSON rules, including fence handling, duplicate-key rejection, and finite-number checks. Every result agrees with its retained parsed object. The primary audit uses these verified parsed objects while preserving recorded parser failures. The audit implementation is exercised by \AuditControlCount{} deterministic controls covering valid receipts, altered balances, holdings and weights, wrong filled-order counts, malformed schemas, invalid nonempty responses, unavailable calls, missing, future, stale, crossed, or wide quotes, and no-imputation/gap-breaking figure behavior. All controls pass. These controls validate implementation behavior; they are not additional model experiments or measurements of semantic detection accuracy.

\paragraph{Analysis policy.}
We report exact counts and descriptive rates. We do not use account-rounds as independent samples for significance testing, annualize a one-week return, or infer a stable model ranking. The experiment was not preregistered. Session-level and activity-conditioned tables make observation imbalance and composition visible. All displayed tables and numerical macros are generated from the normalized audit records.

\section{Results}
\label{sec:results}
\subsection{A valid decision does not ensure an accurate report}
Table~\ref{tab:overview} gives the primary denominators. Cryptocurrency decision validity is \CryptoDecisionPass/\CryptoDecisionDen{} (\CryptoDecisionRate\%), excluding 55 explicitly unavailable decision calls. Structured report accuracy is \CryptoReportPass/\CryptoReportDen{} (\CryptoReportRate\%), excluding 33 API failures and one interrupted report. Using all retained account-rounds as the denominator yields \CryptoDecisionYield\% valid decisions and \CryptoReportYield\% accurate reports. These retained-round yields include unavailable calls in committed rounds but omit attempts that never produced a committed round; they are not all-attempt completion probabilities. Equal-route macro-averages are \CryptoMacroDecisionRate\% for decision validity and \CryptoMacroReportRate\% for report accuracy, compared with the primary pooled rates. Unequal observation counts therefore affect the aggregation, especially for reporting.

\begin{table}[t]
\centering\small
\resizebox{\textwidth}{!}{\begin{tabular}{lrrrrr}
\toprule
Environment & $N$ & Valid decisions & Accurate reports & Kept commitments & Unknown \\
\midrule
Crypto & 7,299 & 7,052/7,244 & 6,597/7,265 & 2,892/3,065 & 2 \\
Equities & 310 & 297/309 & 274/309 & 196/202 & 53 \\
\bottomrule
\end{tabular}
}
\caption{Recomputed primary counts. Decision and report denominators exclude explicitly unavailable calls. Commitment denominators include only nontrivial constraints with eligible marked-state evidence. Unknown refers to declared constraints excluded for unavailable or stale valuation evidence.}
\label{tab:overview}
\end{table}

\begin{figure}[t]
\centering\includegraphics[width=\textwidth]{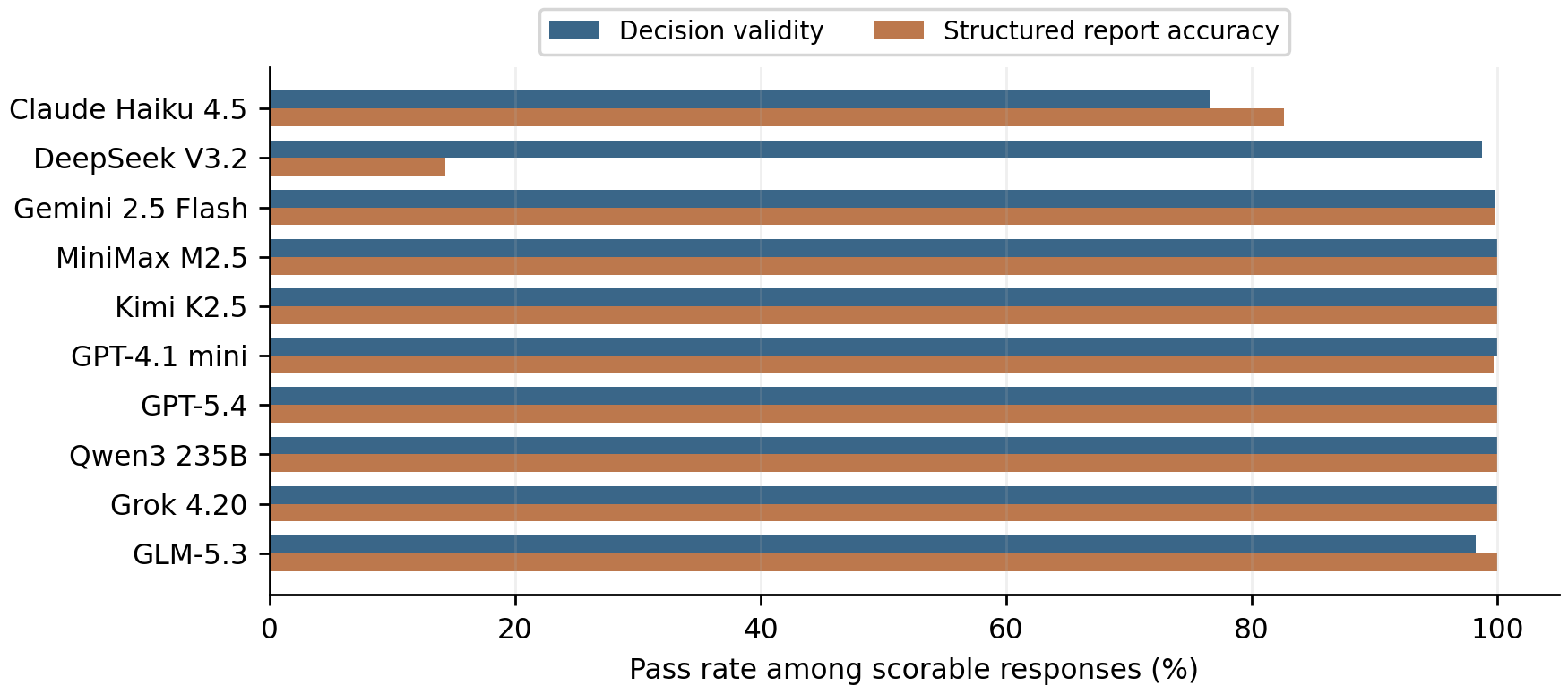}
\caption{Decision validity and structured report accuracy are distinct operational properties. Rates are conditional on scorable responses; unequal sample sizes and infrastructure failures remain visible in the tables.}
\label{fig:rates}
\end{figure}

Differences between requested routes are substantial (Table~\ref{tab:models}, Figure~\ref{fig:rates}). The DeepSeek route produces valid decisions on 620/628 scorable calls, but accurate reports on only 90/628. The Claude route has 556/726 valid decisions and 605/732 accurate reports. All other routes reach 99.73--100\% report accuracy. These are descriptive results under this gateway, prompt, and observation window. They do not establish an immutable ordering of the underlying model families.

\begin{table}[t]
\centering\small\setlength{\tabcolsep}{5pt}
\begin{tabular}{lrrrrrr}
\toprule
Model & $N$ & $V$ (\%) & $A$ (\%) & Cover. (\%) & Adhere. (\%) & Fills \\
\midrule
Claude Haiku 4.5 & 739 & 76.58 & 82.65 & 99.28 & 98.09 & 38 \\
DeepSeek V3.2 & 631 & 98.73 & 14.33 & 0.48 & 100.00 & 9 \\
Gemini 2.5 Flash & 740 & 99.86 & 99.86 & 10.20 & 100.00 & 5 \\
MiniMax M2.5 & 741 & 100.00 & 100.00 & 1.08 & 100.00 & 49 \\
Kimi K2.5 & 741 & 100.00 & 100.00 & 4.19 & 85.25 & 39 \\
GPT-4.1 mini & 742 & 100.00 & 99.73 & 31.30 & 90.04 & 6 \\
GPT-5.4 & 741 & 100.00 & 100.00 & 100.00 & 99.80 & 35 \\
Qwen3 235B & 742 & 100.00 & 100.00 & 0.68 & 100.00 & 3 \\
Grok 4.20 & 741 & 100.00 & 100.00 & 17.49 & 62.89 & 10 \\
GLM-5.3 & 741 & 98.22 & 100.00 & 0.28 & 100.00 & 27 \\
\bottomrule
\end{tabular}

\caption{Cryptocurrency audit by requested model route, combining the settled and partial sessions. $V$ and $A$ use their own scorable denominators. Coverage is the percentage of valid decisions declaring a nontrivial commitment; adherence is per scorable constraint, not per round. Fills are local simulated executions. Exact numerators and denominators appear in Appendix Table~\ref{tab:model-denominators} and the accompanying data.}
\label{tab:models}
\end{table}

The 192 scorable cryptocurrency decision failures comprise 179 recorded JSON-parse errors, seven truncated outputs, and six action-contract violations after parsing. The 668 scorable report failures comprise 395 JSON-parse errors, one truncated output, one empty response, and 271 parsed reports failing the requested fields. Accordingly, poor reporting performance cannot be attributed exclusively to incorrect financial arithmetic: format delivery is itself a large part of the measured failure. Across the 6,868 parsed reports without a recorded response error, 6,597 pass all report checks (96.05\%). This conditional view describes a different, selected population and is not a replacement for the primary accuracy rate.

\subsection{Commitment coverage and activity change the interpretation}
Cryptocurrency agents declare 3,067 nontrivial constraints in 1,772 valid decision rounds: coverage is \CryptoCoverageRate\%. Two constraints have stale or unavailable execution-state evidence and remain unscorable. Of the remaining \CryptoCommitDen{}, \CryptoCommitPass{} pass (\CryptoCommitRate\%). No emitted constraints are excluded as trivial tautologies in this corpus.

\begin{figure}[t]
\centering\includegraphics[width=\textwidth]{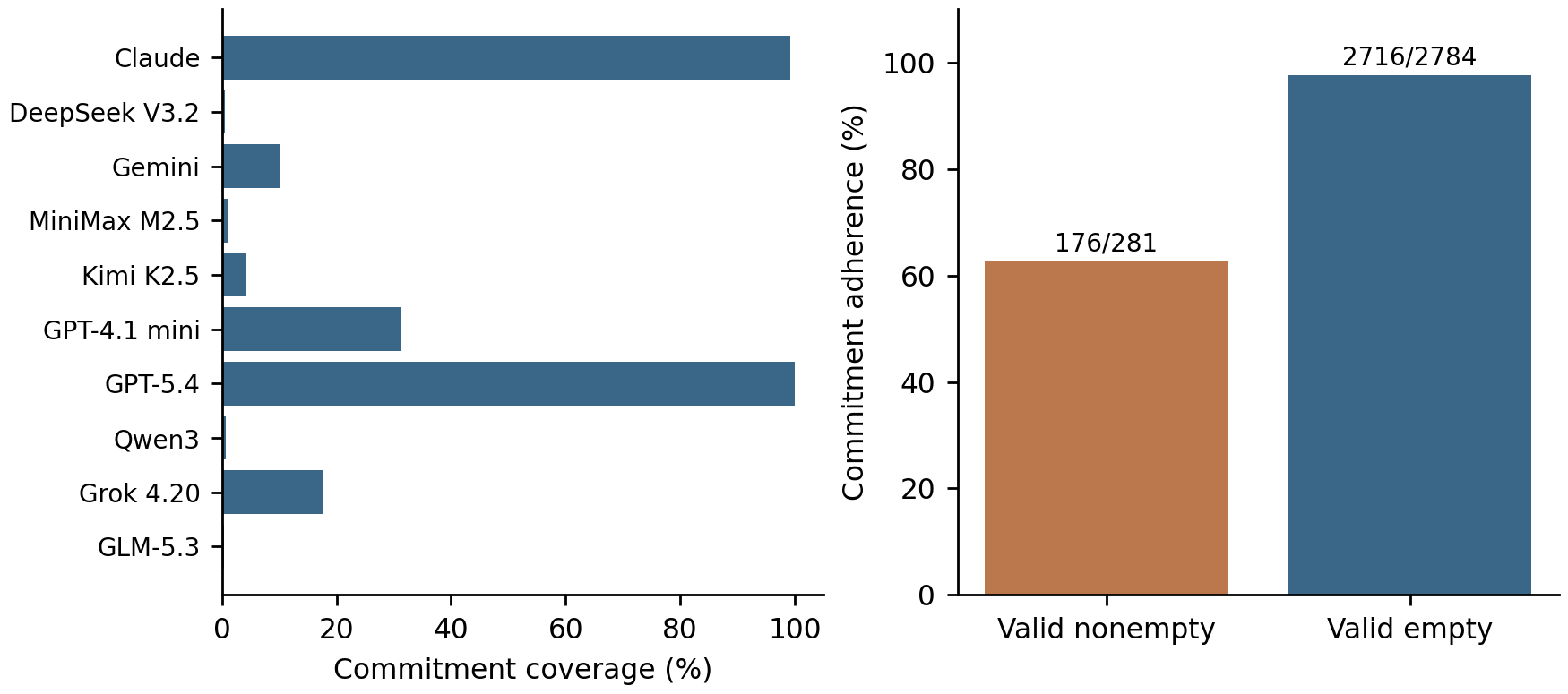}
\caption{Left: declared nontrivial commitment coverage among valid cryptocurrency decisions. Right: compliance among scorable constraints, conditioned on a nonempty or valid empty proposal. Counts above bars are passed/evaluated constraints. Empty proposals dominate the corpus and can conceal materially weaker adherence on active rounds.}
\label{fig:commitments}
\end{figure}

The aggregate conceals two forms of selection. First, commitment coverage varies from 0.28\% for GLM-5.3 to 100\% for GPT-5.4. A 100\% adherence figure based on four GLM constraints or six DeepSeek constraints does not establish the same breadth of evidence as GPT-5.4's 1,007 constraints. Second, activity changes the observed adherence distribution. Only 219/7,052 valid cryptocurrency decisions contain an order (\CryptoActiveRate\%). On those rounds, 176/281 constraints pass (62.63\%); on valid empty proposals, 2,716/2,784 pass (97.56\%). Figure~\ref{fig:commitments} and Table~\ref{tab:strata} expose this difference. It is observational: different models, constraints, and account states populate the strata, so it is not an estimated causal effect of trading.

\begin{table}[t]
\centering\small
\resizebox{\textwidth}{!}{\begin{tabular}{llrrrrr}
\toprule
Env. & Round stratum & $N$ & Reports & \% & Commitments & \% \\
\midrule
Crypto & Nonempty proposal & 219 & 205/219 & 93.61 & 176/281 & 62.63 \\
Crypto & Valid empty proposal & 6833 & 6258/6812 & 91.87 & 2716/2784 & 97.56 \\
Crypto & Any rejection & 87 & 82/87 & 94.25 & 37/131 & 28.24 \\
Equities & Nonempty proposal & 53 & 48/52 & 92.31 & 25/31 & 80.65 \\
Equities & Valid empty proposal & 244 & 221/244 & 90.57 & 171/171 & 100.00 \\
Equities & Any rejection & 25 & 24/25 & 96.00 & 0/0 & -- \\
\bottomrule
\end{tabular}
}
\caption{Activity-conditioned audit. Reports and commitments have different eligibility denominators. Any-rejection strata can overlap other strata; these rows are not a partition to sum. Undefined adherence is shown as a dash. An empty valid proposal is distinct from a malformed or unavailable decision.}
\label{tab:strata}
\end{table}

Commitment violations need not imply a new unsafe trade. In the rejected-proposal stratum, only 37/131 commitments pass: the final portfolio can remain unchanged after a gate rejection while failing a promise about the intended post-execution allocation. Conversely, an empty proposal can preserve a compliant portfolio. This is why we check explicit promises against the actual recorded state, while separately retaining the proposed orders and gate outcome.

\subsection{Execution controls and ledger fidelity are different measurements}
The cryptocurrency traces contain \CryptoOrders{} orders in contract-valid proposals, of which \CryptoFills{} fill and 117 are rejected. All recorded rejections carry the gate's \texttt{quality\_rejected} reason. The gate fails on 1,094 account-rounds: 760 valid empty proposals, 87 valid nonempty proposals, and 247 invalid or unavailable decision responses (Table~\ref{tab:gate-populations}). The last category must not be interpreted as an empty valid decision. Four parsed invalid cryptocurrency responses contain 13 recognizable order objects, excluded from admitted-order totals because the full decision violates the action contract. Unparseable responses may contain further uninterpretable intent. The failed-gate count is therefore neither a rejected-order count nor a count of 1,094 attempted trades.

\begin{table}[t]
\centering\small\begin{tabular}{lrrr}
\toprule
Decision category & Account-rounds & Gate pass & Gate fail \\
\midrule
Valid nonempty & 219 & 132 & 87 \\
Valid empty & 6833 & 6073 & 760 \\
Invalid or unavailable & 247 & 0 & 247 \\
\bottomrule
\end{tabular}

\caption{Cryptocurrency gate populations. Contract-invalid or unavailable responses form a third input category; zero admitted orders in that category does not establish a valid choice to do nothing. Counts partition retained account-rounds.}
\label{tab:gate-populations}
\end{table}

Gate checks are not mutually exclusive. Recorded failures include 594 maximum-position checks, 252 cash-floor checks, 252 total-exposure checks, five funds-and-positions checks, and one fresh-quote check, alongside 247 invalid or unavailable decision-schema cases. Within valid empty proposals alone, failed checks comprise 524 maximum-position, 235 cash-floor, 235 total-exposure, and one fresh-quote check, with overlap among risk checks. Risk-limit failures in this subgroup concern an existing marked state without a newly submitted order; they are distinct from quote unavailability and invalid input. Existing holdings can cross a limit as prices change. The gate also changes which actions become fills, preventing an attribution of the observed portfolio path solely to the language model.

Every retained account-round reconciles cash and positions against recorded filled orders: 7,299/7,299 in cryptocurrency and 310/310 in equities. Equity arithmetic has complete execution-price evidence on 279/310 rounds, all of which pass; the remaining 31 have no complete mark evidence and receive no equity-arithmetic score. Cryptocurrency equity arithmetic passes on all 7,299 recorded states, including a stale marked state whose numerical equality is valid but whose two commitments are conservatively unscorable. Internal arithmetic agreement and current market validity are separate checks.

\subsection{The equity pilot illustrates environment-error attribution}
The smaller equity corpus yields \EquityDecisionPass/\EquityDecisionDen{} valid decisions (\EquityDecisionRate\%) and \EquityReportPass/\EquityReportDen{} accurate reports (\EquityReportRate\%). One decision call and one report call are explicitly unavailable. Of 168 orders in contract-valid proposals, 63 fill, 100 receive \texttt{market\_unavailable}, and five are rejected for insufficient cash. One additional invalid parsed response contains three recognizable order objects and is excluded from the admitted-order population. Reporting remains independently evaluable from the supplied receipt: 24/25 scorable reports on rounds containing a rejection are accurate.

Agents declare 255 nontrivial constraints in 170 valid rounds, but 53 lack eligible execution-state evidence. Scoring those against carried-forward balances would turn unavailable evidence into apparent success. With those checks excluded, 196/202 constraints pass (\EquityCommitRate\%). The active-proposal stratum has 25/31 passed constraints, compared with 171/171 on valid empty proposals. These counts echo the importance of activity conditioning, but the different windows and controls do not support a cross-market generalization claim.

\subsection{A concrete factual discrepancy}
One retained DeepSeek report on 2026-10-03 04:42:49 UTC correctly reproduces cash of \$8,201.03107724632, equity of \$9,985.95607724632, and a position of 15 SOL. Its summary states that no orders were submitted or executed. The recorded order list is empty, yet \texttt{filled\_orders} equals two. The report therefore fails exactly the round-specific filled-order count while passing cash, equity, positions, and schema. A second record returns 1,584,655,368 as that count on another empty-order round. These cases demonstrate the value of field-level reconciliation; they do not identify a mechanism for the error or intentional misrepresentation. Case identifiers and source digests are supplied in the reproducibility package.

\subsection{Accurate reporting and short-window returns answer different questions}
For a common marked outcome, we use the settled cryptocurrency session alone. Each account's latest eligible saved valuation is at 2026-10-07 03:42:24 UTC, approximately five seconds before the recorded deadline. We independently reconstruct all ten endpoint equities from the last saved cash and holdings and one common cached quote bundle, whose recorded quote and receipt times precede the deadline; the maximum absolute discrepancy is below $2\times10^{-12}$. The source artifact carrying that cached bundle was written after the deadline, so this is consistency with recorded pre-deadline quote metadata rather than independent proof of a contemporaneous commitment. Table~\ref{tab:performance} reports full-window marked returns from the recorded initial allocation, after recorded fees and without forced liquidation. All ten returns are negative, ranging from $-1.448\%$ to $-0.157\%$. Seven routes have 100\% conditional reporting accuracy during this settled session, yet each loses value over the same window. A correct account report is therefore not sufficient evidence of a positive return in these observations.

\begin{figure}[t]
\centering\includegraphics[width=\textwidth]{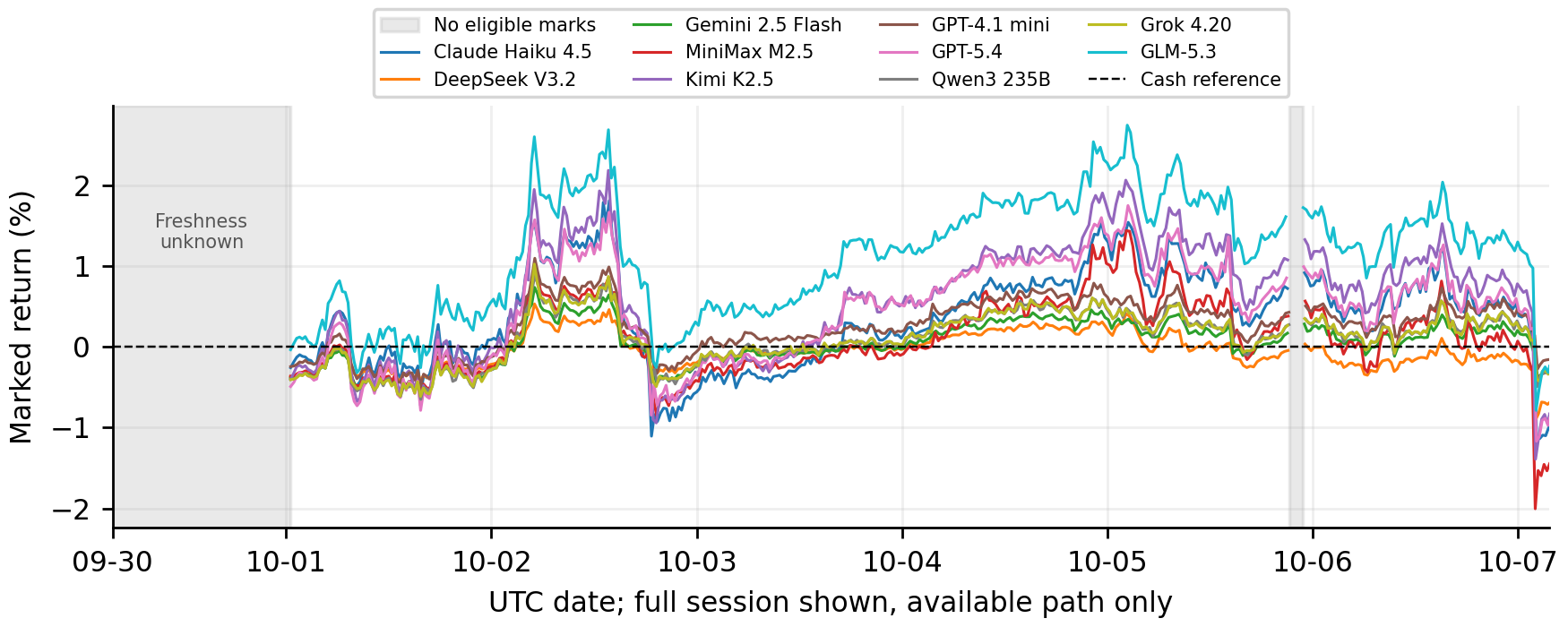}
\caption{Available explicitly nonstale saved marks, with returns measured from initial cash. The shaded initial 20-hour-49-minute interval has no eligible marks; the later October 5 gap is approximately 97 minutes. Curves are subsampled and broken across source gaps longer than five minutes, without imputing missing observations. The full-session endpoint is retained. The cash reference assumes no interest or trading and is not an exposure-matched market benchmark.}
\label{fig:returns}
\end{figure}

Figure~\ref{fig:returns} shows a narrower path: only saved valuations explicitly flagged nonstale. The first eligible point is 2026-10-01 00:31:24 UTC, about 20 hours 49 minutes after the session starts. Earlier marks have unknown staleness and are excluded, not imputed. A later saved-mark gap on October 5 spans approximately 97 minutes. The figure shades the initial exclusion and breaks curves across source gaps longer than five minutes. Endpoint returns remain defined by initial cash and final eligible state; an incomplete observed path must not be presented as a complete seven-day valuation history.

\begin{table}[t]
\centering\small\begin{tabular}{lrrrr}
\toprule
Model & $N$ & Return (\%) & $A$ (\%) & Cash (\%) \\
\midrule
Claude Haiku 4.5 & 664 & -0.970 & 80.67 & 27.4 \\
DeepSeek V3.2 & 615 & -0.691 & 14.38 & 71.4 \\
Gemini 2.5 Flash & 665 & -0.313 & 99.85 & 75.2 \\
MiniMax M2.5 & 666 & -1.448 & 100.00 & 20.8 \\
Kimi K2.5 & 666 & -0.833 & 100.00 & 20.3 \\
GPT-4.1 mini & 667 & -0.157 & 100.00 & 70.4 \\
GPT-5.4 & 666 & -0.886 & 100.00 & 27.2 \\
Qwen3 235B & 667 & -0.315 & 100.00 & 70.5 \\
Grok 4.20 & 666 & -0.314 & 100.00 & 70.5 \\
GLM-5.3 & 666 & -0.241 & 100.00 & 21.4 \\
\bottomrule
\end{tabular}

\caption{Settled seven-day cryptocurrency session only. $N$ is retained account-rounds; $A$ is report accuracy conditional on a scorable response. Return uses the same saved endpoint for all accounts. Cash is the final fraction of marked equity. These are descriptive marked outcomes, not estimates of alpha.}
\label{tab:performance}
\end{table}

We do not infer that concordance is uncorrelated with returns. Ten dependent portfolio paths over one week cannot establish that conclusion. Nor do we benchmark investment skill against a passive market portfolio here. The observation supports the narrower distinction already built into the protocol: account-reporting correctness and investment outcomes should remain separate evaluation axes.

\subsection{Recorded resource use}
The cryptocurrency call table contains 14,833 lifecycle entries, compared with 14,598 decision/report slots in committed rounds. The difference reflects a broader population of attempts and recovery records; it is not evidence of 235 successful additional decisions. Prompt and completion counts are both present on 14,737 entries. Their recorded sums are 202,982,942 prompt tokens and 8,341,624 completion tokens, or \TokenTotal{} tokens in total. Missing usage is not imputed as measured zero. Table~\ref{tab:tokens} provides route-level accounting. These totals describe the recorded workload, including attempts outside committed rounds, rather than a cost-normalized model comparison. We report no reconstructed dollar bill or claimed savings from an untested routing policy.

\section{Limitations}
\label{sec:limitations}

\paragraph{Observational and dependent evidence.}
This is a retrospective audit of retained forward-running traces, not a preregistered controlled experiment. The two environments cover different assets, dates, schedules, and execution policies. Consecutive account-rounds share market conditions, portfolio history, and often input data; their counts must not be read as independent replications. We have neither randomized ablations of audit feedback nor repeated independent runs sufficient to isolate model effects from market path and accumulated state. Differences between models are descriptive of these traces. The short windows do not establish stable rankings, long-run alpha, or generalization across market regimes.

\paragraph{The evidence boundary.}
Deterministic checks avoid requiring a semantic judge, at the cost of a narrower claim. Decision validity measures format and allowed values; reporting accuracy measures structured factual fields. Neither establishes that the rationale matches the orders or that narrative explanations reveal the model's internal reasoning. Optional commitment compliance concerns the constraints actually emitted and evaluable; coverage is essential, and selection into commitment-making is uncontrolled. No human-labeled semantic benchmark supports a broader faithfulness or deception claim. Likewise, accurate reporting after seeing a receipt does not validate an ex-ante forecast.

\paragraph{Policy and execution confounding.}
The cryptocurrency agents receive algorithmic analysis, including momentum signals, and operate behind a deterministic proposal gate. These are components of the evaluated system. They can affect activity, exposure, returns, and apparent adherence; this study cannot attribute those outcomes solely to the language model. Whole-proposal rejection can suppress feasible orders alongside infeasible ones. Observation-to-execution latency and scheduling vary, and the retained traces do not provide a synchronized randomized comparison of execution times. Equity and cryptocurrency results therefore remain separate rather than being interpreted as a clean cross-market transfer test.

\paragraph{Local accounting and provenance.}
All fills are generated by a local paper executor using recorded public-market quotes. The audit verifies arithmetic against retained fills, not whether an exchange would have executed them. It does not model queue priority, partial fills, depth-dependent impact, or all venue-specific costs. Ledger agreement can also hold with stale prices and cannot certify external market truth. Saved valuation freshness is unknown for the initial 20 hours 49 minutes, and the later explicitly nonstale path has a roughly 97-minute gap; the displayed curve is therefore incomplete even though common endpoint values can be reconstructed. Recorded pre-deadline quote metadata is not independently notarized evidence of when it became available. One recovered cryptocurrency record has a nonmonotonic round ordinal; chronological state continuity is restored by timestamp ordering, and the record is retained with the anomaly disclosed. Hash agreement detects changes relative to the stored artifact index, not errors already present when artifacts were written. Missing ancillary execution-attempt artifacts limit complete reconstruction of attempts outside committed round records, so retained-round yield cannot estimate success across all started attempts.

\paragraph{Model and cost attribution.}
Model names identify requested inference routes. Returned identifiers and call metadata improve traceability but do not independently attest immutable model weights, provider routing, or serving configuration. Prompt and operational differences also constrain attribution. Token totals describe observed workload, not a normalized computational budget or verified monetary bill. More complete identity attestation, prospective protocol freezing, semantic annotations, and controlled repeated trials are needed before broader causal or financial conclusions.

\section{Discussion}
The central implication is a separation of evidential claims. Action-contract validity establishes that an executor can interpret a decision. Commitment adherence establishes that particular stated constraints hold in a recorded state. Ledger reconciliation establishes arithmetic agreement with fills. Structured reporting accuracy establishes that selected factual claims match a supplied receipt. None alone establishes the other properties, and none establishes profitable investment or transparent internal reasoning.

The data also show why denominators belong in the result, rather than only in an appendix. An agent that rarely declares a constraint has few opportunities to violate one. A system that usually does nothing has many straightforward account reports. A gate that blocks a whole proposal changes the portfolio against which promises are checked. Combining these outcomes into one score without coverage, activity, and eligibility can reward the composition of the observations more than the intended behavior. We therefore provide a profile of measurements and retain undefined outcomes explicitly.

The audit remains useful even when a failure is elementary. A malformed JSON report may be less intellectually interesting than an elaborate false rationale, but it still prevents a downstream application from obtaining the contracted factual output. Conversely, detecting such a failure does not establish that the model cannot reason about finance. Reporting parse, schema, field, and infrastructure outcomes separately constrains this interpretation. The same discipline applies to execution: a missing market quote should not be counted as an invented trade, and a correct report of a rejected order should not be penalized merely because no trade occurred.

Three extensions follow from these limits. First, a prospectively frozen protocol with repeated independent accounts and explicit model-identity records would make comparisons less dependent on the observed path and recovery history. Second, human-annotated rationale--order and summary--receipt pairs would extend the structured checks to semantic concordance; independent adjudication would be needed to avoid treating one language-model judge as an oracle. Third, randomized feedback and gate ablations could test whether audits improve subsequent behavior or merely change activity and exposure. These are future experiments, not effects demonstrated by the present corpus.

\section{Conclusion}
FIDES-Live extends the FIDES measurement perspective from generated trading strategies to sequential agent traces. Its audit distinguishes valid decisions, optional commitments, local execution, account-state arithmetic, and post-execution factual reports while making unavailable evidence visible. Across \AllRounds{} retained account-rounds, ledger reconciliation coexists with report failures, optional commitment coverage varies widely, and adherence differs substantially between active and empty proposals. The results support artifact-level auditing as a complement to financial outcome reporting. They establish neither persistent model rankings nor profitable trading; they establish a reproducible account of which observable claims agree with the recorded actions and states, and where that evidence stops.

\paragraph{Data and code availability.}
The accompanying source package includes the generated tables, figures, normalized audit outcomes, source digests, and offline analysis code. Original local-market and inference artifacts are retained for provenance checking and are not assumed to be publicly hosted. No credentials or environment files are included. The normalized outcomes reproduce the reported audit aggregates; regeneration from raw evidence requires access to the retained source corpus. No real-money orders were placed for this study.

\bibliographystyle{plainnat}
\bibliography{references}

@misc{tian2026fides,
  title = {{FIDES}: A Concordance Protocol for {LLM}-Generated Trading Strategies},
  author = {Tian, Arther and Ding, Alex and Wu, Simon and Chan, Aaron},
  year = {2026},
  eprint = {2608.23308},
  archivePrefix = {arXiv},
  primaryClass = {cs.CR},
  doi = {10.48550/arXiv.2608.23308},
  url = {https://arxiv.org/abs/2608.23308}
}

@misc{xiao2024tradingagents,
  title = {{TradingAgents}: Multi-Agents {LLM} Financial Trading Framework},
  author = {Xiao, Yijia and Sun, Edward and Luo, Di and Wang, Wei},
  year = {2024},
  eprint = {2412.20138},
  archivePrefix = {arXiv},
  doi = {10.48550/arXiv.2412.20138},
  url = {https://arxiv.org/abs/2412.20138}
}

@misc{chen2025stockbench,
  title = {{StockBench}: Can {LLM} Agents Trade Stocks Profitably In Real-world Markets?},
  author = {Chen, Yanxu and Yao, Zijun and Liu, Yantao and Xin, Amy and Ye, Jin and Yu, Jianing and Hou, Lei and Li, Juanzi},
  year = {2025},
  eprint = {2510.02209},
  archivePrefix = {arXiv},
  doi = {10.48550/arXiv.2510.02209},
  url = {https://arxiv.org/abs/2510.02209}
}

@misc{qian2025whenagents,
  title = {When Agents Trade: Live Multi-Market Trading Benchmark for {LLM} Agents},
  author = {Qian, Lingfei and Peng, Xueqing and Wang, Yan and Zhang, Vincent Jim and He, Huan and Smith, Hanley and Han, Yi and He, Yueru and Li, Haohang and Cao, Yupeng and Yu, Yangyang and Lopez-Lira, Alejandro and Lu, Peng and Nie, Jian-Yun and Xiong, Guojun and Huang, Jimin and Ananiadou, Sophia},
  year = {2025},
  eprint = {2510.11695},
  archivePrefix = {arXiv},
  doi = {10.48550/arXiv.2510.11695},
  url = {https://arxiv.org/abs/2510.11695}
}

@misc{yu2025livetradebench,
  title = {{LiveTradeBench}: Seeking Real-World Alpha with Large Language Models},
  author = {Yu, Haofei and Li, Fenghai and You, Jiaxuan},
  year = {2025},
  eprint = {2511.03628},
  archivePrefix = {arXiv},
  primaryClass = {q-fin.TR},
  doi = {10.48550/arXiv.2511.03628},
  url = {https://arxiv.org/abs/2511.03628}
}

@misc{zhang2026alphaforgebench,
  title = {{AlphaForgeBench}: Benchmarking End-to-End Trading Strategy Design with Large Language Models},
  author = {Zhang, Wentao and Zhao, Mingxuan and Gao, Jincheng and You, Jieshun and Jia, Huaiyu and Zhao, Yilei and An, Bo and Sun, Shuo},
  year = {2026},
  eprint = {2602.18481},
  archivePrefix = {arXiv},
  primaryClass = {q-fin.TR},
  doi = {10.48550/arXiv.2602.18481},
  url = {https://arxiv.org/abs/2602.18481}
}

@misc{liu2023agentbench,
  title = {{AgentBench}: Evaluating {LLM}s as Agents},
  author = {Liu, Xiao and Yu, Hao and Zhang, Hanchen and Xu, Yifan and Lei, Xuanyu and Lai, Hanyu and Gu, Yu and Ding, Hangliang and Men, Kaiwen and Yang, Kejuan and Zhang, Shudan and Deng, Xiang and Zeng, Aohan and Du, Zhengxiao and Zhang, Chenhui and Shen, Sheng and Zhang, Tianjun and Su, Yu and Sun, Huan and Huang, Minlie and Dong, Yuxiao and Tang, Jie},
  year = {2023},
  eprint = {2308.03688},
  archivePrefix = {arXiv},
  doi = {10.48550/arXiv.2308.03688},
  url = {https://arxiv.org/abs/2308.03688},
  note = {Published at ICLR 2024}
}

@misc{yao2024taubench,
  title = {{$\tau$}-bench: A Benchmark for Tool-Agent-User Interaction in Real-World Domains},
  author = {Yao, Shunyu and Shinn, Noah and Razavi, Pedram and Narasimhan, Karthik},
  year = {2024},
  eprint = {2406.12045},
  archivePrefix = {arXiv},
  doi = {10.48550/arXiv.2406.12045},
  url = {https://arxiv.org/abs/2406.12045}
}

@inproceedings{turpin2023unfaithful,
  title = {Language Models Don't Always Say What They Think: Unfaithful Explanations in Chain-of-Thought Prompting},
  author = {Turpin, Miles and Michael, Julian and Perez, Ethan and Bowman, Samuel R.},
  booktitle = {Advances in Neural Information Processing Systems},
  volume = {36},
  year = {2023},
  url = {https://arxiv.org/abs/2305.04388}
}

@misc{tian2026robustpoq,
  title = {Adaptive and Robust Cost-Aware Proof of Quality for Decentralized {LLM} Inference Networks},
  author = {Tian, Arther and Ding, Alex and Chen, Frank and Wu, Simon and Chan, Aaron},
  year = {2026},
  eprint = {2601.21189},
  archivePrefix = {arXiv},
  url = {https://arxiv.org/abs/2601.21189}
}

@misc{tian2026multidimensional,
  title = {A Multi-Dimensional Quality Scoring Framework for Decentralized {LLM} Inference with Proof of Quality},
  author = {Tian, Arther and Ding, Alex and Chen, Frank and Wu, Simon and Chan, Aaron},
  year = {2026},
  eprint = {2603.04028},
  archivePrefix = {arXiv},
  url = {https://arxiv.org/abs/2603.04028}
}

@misc{tian2026poqjudge,
  title = {{PoQ-Judge}: A Multi-Architecture Evaluation Framework for Cost-Aware Proof-of-Quality in Decentralized {LLM} Inference},
  author = {Tian, Arther and Ding, Alex and Chen, Frank and Wu, Simon and Chan, Aaron},
  year = {2026},
  eprint = {2606.11196},
  archivePrefix = {arXiv},
  url = {https://arxiv.org/abs/2606.11196}
}

@misc{tian2026sfga,
  title = {{SFGA}: A Statistics-First Gating Architecture with Adjudicative Escalation for Trustworthy {SFT} Data Procurement},
  author = {Tian, Arther and Ding, Alex and Chen, Frank and Wu, Simon and Chan, Aaron},
  year = {2026},
  eprint = {2607.18960},
  archivePrefix = {arXiv},
  url = {https://arxiv.org/abs/2607.18960}
}
\clearpage
\appendix
\section{Recorded Routes and Protocol Differences}
\begin{table}[h]
\centering\small
\begin{tabular}{ll}
\toprule
Abbreviation & Exact requested route\\
\midrule
GPT-4.1 mini & \texttt{openai/gpt-4.1-mini}\\
GPT-5.4 & \texttt{openai/gpt-5.4}\\
DeepSeek V3.2 & \texttt{deepseek/deepseek-v3.2}\\
Gemini 2.5 Flash & \texttt{google/gemini-2.5-flash}\\
Claude Haiku 4.5 & \texttt{anthropic/claude-haiku-4.5}\\
Qwen3 235B & \texttt{qwen/qwen3-235b-a22b-instruct-2507}\\
Kimi K2.5 & \texttt{moonshotai/kimi-k2.5}\\
GLM-5.3 & \texttt{z-ai/glm-5.3}\\
MiniMax M2.5 & \texttt{minimax/minimax-m2.5}\\
Grok 4.20 & \texttt{x-ai/grok-4.20-non-reasoning}\\
\bottomrule
\end{tabular}
\caption{Model names denote request routes, not independently attested immutable model weights.}
\label{tab:routes}
\end{table}

\begin{table}[h]
\centering\small
\begin{tabular}{>{\raggedright\arraybackslash}p{4.0cm}>{\raggedright\arraybackslash}p{4.2cm}>{\raggedright\arraybackslash}p{4.7cm}}
\toprule
Setting & Cryptocurrency & Equity pilot\\
\midrule
Assets & Six spot assets & Eight ETFs initially; fifteen assets in main run\\
Decision interval & 15 minutes & 5 minutes initially; 15 minutes in main run\\
Initial allocation & \$10,000 per model/session & \$10,000 per model/run\\
Historical context & Up to 90 daily bars & Up to 30 daily bars\\
Fee / adverse slippage & 10 / 2 basis points & 1 / 2 basis points\\
Cash / position / exposure & 20\% / 30\% / 80\% fixed gate & No matching balanced gate\\
Advisory analysis & Deterministic momentum/volatility & None supplied\\
Maximum orders & 12 & 20\\
Quote age / spread & 90 seconds / 100 basis points & 120 seconds / 100 basis points in main run\\
Execution synchronization & Per-account snapshots & Joint decision-batch barrier\\
\bottomrule
\end{tabular}
\caption{Material recorded differences. These conditions are not pooled into a causal cross-market comparison.}
\label{tab:settings}
\end{table}

The cryptocurrency decision system prompt is present in all 7,299 round artifacts with one SHA-256 value; 7,298 report prompts share one report hash, with one interrupted report lacking a prompt. Self-hashes match where present. The decision hash is \hash{a6eb0fdbf2b7569db5db457f5edd5e907960f0e7509223d1404d2314d153cfd6}; the report hash is \hash{2ab3a395a589d6a35631683867ebb5cef84b4f73fb4553d952789e8f558871c3}. Long hashes in the source data provide exact byte-level identification. Two wordings of an advisory strategy-role field occur in 5,505 and 1,794 requests, respectively; recorded numerical risk and execution settings match. Prompt stability therefore does not imply that every complete request payload is identical.

\clearpage
\section{Resource Accounting and Daily Observation Counts}
\begin{table}[H]
\centering\small\begin{tabular}{lrrrr}
\toprule
Model & Calls & Tokens (M) & Usage pairs & Median seconds \\
\midrule
Claude Haiku 4.5 & 1484 & 27.54 & 1463 & 9.43 \\
DeepSeek V3.2 & 1484 & 16.52 & 1476 & 5.48 \\
Gemini 2.5 Flash & 1484 & 28.83 & 1476 & 2.47 \\
MiniMax M2.5 & 1483 & 17.31 & 1479 & 10.77 \\
Kimi K2.5 & 1483 & 22.50 & 1480 & 11.68 \\
GPT-4.1 mini & 1484 & 16.02 & 1479 & 2.40 \\
GPT-5.4 & 1482 & 16.47 & 1476 & 4.55 \\
Qwen3 235B & 1484 & 25.29 & 1476 & 8.14 \\
Grok 4.20 & 1482 & 16.17 & 1468 & 2.29 \\
GLM-5.3 & 1483 & 24.67 & 1464 & 21.52 \\
\bottomrule
\end{tabular}

\caption{Cryptocurrency call-table accounting, including attempts outside committed rounds. Tokens are recorded prompt plus completion counts; missing usage contributes no measured amount. Usage pairs counts calls with both components available. Median elapsed seconds uses recorded values and is descriptive of the gateway/system, not a hardware-normalized speed benchmark.}
\label{tab:tokens}
\end{table}
\begin{table}[H]
\centering\small
\begin{tabular}{lrrr}
\toprule
UTC date & Account-rounds & Accurate/scorable reports & Accuracy (\%) \\
\midrule
2026-09-30 & 820 & 686/809 & 84.80 \\
2026-10-01 & 959 & 805/952 & 84.56 \\
2026-10-02 & 960 & 860/960 & 89.58 \\
2026-10-03 & 959 & 872/957 & 91.12 \\
2026-10-04 & 957 & 867/949 & 91.36 \\
2026-10-05 & 906 & 814/900 & 90.44 \\
2026-10-06 & 913 & 890/913 & 97.48 \\
2026-10-07 & 825 & 803/825 & 97.33 \\
\bottomrule
\end{tabular}

\caption{Cryptocurrency observation counts by UTC date. Partial first/last days and operational gaps remain visible. Accurate/scorable reports use the same primary definition as the main text.}
\label{tab:daily}
\end{table}

\subsection*{Exact model-level denominators}
\begin{table}[H]
\centering\small\setlength{\tabcolsep}{4pt}
\begin{tabular}{lrrrr}
\toprule
Route & Valid/scorable & Accurate/scorable & Declaring/valid & Passed/evaluable \\
\midrule
Claude Haiku 4.5 & 556/726 & 605/732 & 552/556 & 1081/1102 \\
DeepSeek V3.2 & 620/628 & 90/628 & 3/620 & 6/6 \\
Gemini 2.5 Flash & 735/736 & 738/739 & 75/735 & 150/150 \\
MiniMax M2.5 & 740/740 & 738/738 & 8/740 & 8/8 \\
Kimi K2.5 & 739/739 & 740/740 & 31/739 & 52/61 \\
GPT-4.1 mini & 738/738 & 739/741 & 231/738 & 416/462 \\
GPT-5.4 & 737/737 & 739/739 & 737/737 & 1005/1007 \\
Qwen3 235B & 738/738 & 738/738 & 5/738 & 9/9 \\
Grok 4.20 & 732/732 & 736/736 & 128/732 & 161/256 \\
GLM-5.3 & 717/730 & 734/734 & 2/717 & 4/4 \\
\bottomrule
\end{tabular}

\caption{Cryptocurrency exact numerators and denominators corresponding to Table~\ref{tab:models}. Declaring/valid is commitment coverage; passed/evaluable counts individual nontrivial constraints. A perfect rate based on a few constraints is not evidence of broad coverage.}
\label{tab:model-denominators}
\end{table}

\section{Audit Reconstruction and Case Traceability}
\paragraph{Inclusion.}
The corpus includes every available populated original equity paper run and every indexed cryptocurrency round belonging to API-driven arena paper accounts at the audit snapshot. Empty initializations, deterministic demonstrations, historical replay, copied equity archives, and personal fixtures are excluded. Combining datasets refers to a common audit schema, not to relabeling their actual market, policy, or source.

\paragraph{Missingness.}
The 840 unavailable ancillary attempt artifacts are counted as missing provenance, while the corresponding round and call populations are assessed from their independently checked files. One orphan started call artifact is not counted as a completed response. Database compaction is not treated as data loss when the full original is available. Report-accuracy component failures include unparseable or absent factual fields; they should not be read as counts of proven numerical fabrication.

\paragraph{Chronology.}
The recovered MiniMax observation has record identifier \hash{ce8e50adf8944db8a4e777ce0b64c598}, stored round ordinal 84, and timestamp 2026-09-30 06:57:53 UTC. Its original request payload identifies round 14. Timestamp ordering places it back in its account sequence and restores continuity. We retain both identifiers and the source digest; no original evidence is overwritten.

\paragraph{Reporting cases.}
The empty-order DeepSeek example with reported filled count two has record identifier \hash{04a5af03a4d447f8abc429367025fa72} and source SHA-256 \hash{8c49dceef81e8665158786330c4cff942163cbbd8ec4ebec0b6d41b6d454164a}. The example with filled count 1,584,655,368 has identifier \hash{01289cb3d9944bfea866112cdf8e5e57} and digest \hash{3e3a0bca1b239b993f529c525eff737d9454b6bf0d60b56e0ad72636e19fc279}. These selected cases illustrate already-defined deterministic failures and are not a human-labeled semantic test set.

\paragraph{Reproduction.}
The offline audit validates source identity, reconstructs each account-round into a common schema, computes content eligibility, rechecks action contracts and report fields, and reconciles account transitions. The summarization stage generates all tables, numerical macros, and figures. Normalized outcomes carry opaque record identifiers and source digests, allowing aggregate reproduction without exposing machine paths or credentials. Recomputing all factual checks from first principles additionally requires the original account and response artifacts. The source README distinguishes these two reproduction levels.

\section{Historical Output Contracts}
The following text is recovered from a retained historical cryptocurrency
request, rather than reconstructed from the current application code.
Decision excerpts omit other instructions; the report prompt is reproduced
in full. The original record identifier, source digest, and full-system-prompt
hashes accompany the source package. These contracts explain why a parsed
but structurally invalid decision or a cumulative rather than round-specific
fill count fails the requested interface.

\paragraph{Decision contract (excerpt).}
\begingroup\footnotesize
\begin{verbatim}
Return ONLY a JSON object. Required fields:
- rationale: a nonempty string explaining your own plan, evidence, sizing and risks.
- orders: an array of order objects. Each order has exactly symbol (an allowed
  ticker string), side ("buy" or "sell"), and quantity (a positive asset-unit count).
Optional field: commitments, an object containing any numerical promises you make.
\end{verbatim}
\endgroup
\paragraph{Numerical commitments (excerpt).}
\begingroup\footnotesize
\begin{verbatim}
commitments is optional (or {}). Only include numerical promises you actually
make about the portfolio IMMEDIATELY after execution, marked at execution mids.
They are evaluated, not imposed on you. They do not describe future returns.
Supported promises are min_cash_ratio and max_position_ratio, numbers in [0,1].
Do not add promises merely to fill the format. Cash >= 0 and position <= 100%
already follow from the trading rules and are not informative promises.
\end{verbatim}
\endgroup
\paragraph{Post-execution report (full system prompt).}
\begingroup\footnotesize
\begin{verbatim}
Report what ACTUALLY happened in this paper trading round, using only
the supplied execution receipt. Distinguish rejected orders from filled orders.
Return ONLY JSON with exactly these fields: summary (a nonempty factual recap),
cash (number), equity (number), positions (ticker-to-share-count object),
filled_orders (integer count of filled orders in this round).
positions contains share quantities, not dollar amounts. Include all nonzero
holdings. Do not invent successful trades or profitability. This is a report,
not another chance to place orders.
Copy the supplied account values faithfully: money within 0.01 and share counts
within 0.000001. Preserve the receipt's precision rather than rounding holdings.
\end{verbatim}
\endgroup

\end{document}